\documentclass[conference]{IEEEtran} 
\usepackage{graphicx} 
\usepackage{cite}
\usepackage{multirow}
\usepackage[hidelinks]{hyperref}

\title{From Regulation to Requirements: An Automated Requirement Derivation and Explanation Pipeline}
\author{
Pavithra PM Nair and Preethu Rose Anish\\
Tata Consultancy Services\\
Email: \{pavithra.nair, preethu.rose\}@tcs.com
}

\begin{document}

\maketitle

\begin{abstract}
Ensuring software compliance with regulations such as the General Data Protection Regulation (GDPR) and the Artificial Intelligence Act (EU AI Act) poses a significant challenge, as requirements engineers must translate complex legal text into actionable software requirements — a process that remains largely manual and error-prone in practice. We present an automated regulation-to-requirements pipeline that identifies requirement-bearing clauses in regulatory documents and derives system-agnostic software requirements, accompanied by plain-language explanations, traceable to their legal sources. We evaluate the pipeline on the full clause sets of the GDPR (398 clauses) and the EU AI Act (574 clauses). For requirement-bearing clause identification, the approach achieves macro-averaged F1 scores of 0.82 and 0.78, respectively, outperforming a SetFit-based baseline. Human evaluation shows high completeness (4.60 and 4.45) and correctness (3.74 and 3.54) of derived requirements, while explanation clarity scores are near-ceiling (4.92 and 4.94) on a 1--5 scale. We implement the approach in Reg2Req, a publicly released tool that further supports requirement classification, use case seeding, cross-reference analysis, definition indexing, and a traceability matrix to operationalize regulatory compliance in practice. A user study with 25 practitioners shows that the plain-language explanations significantly improve comprehension of derived requirements and confidence in acting on them ($p < 0.001$), and that all participants would use Reg2Req as a starting point for deriving software requirements from a regulation.
\end{abstract}

\begin{IEEEkeywords}
requirements engineering, regulatory compliance, natural language processing, large language models
\end{IEEEkeywords}

\section{Introduction}
Modern software systems increasingly operate under stringent regulatory frameworks~\cite{kosenkov2025systematic, abualhaija2024ai}. Data protection law, AI regulation, healthcare standards, and sector-specific legislation all impose mandatory requirements on how systems must be designed, how they must behave at runtime, and what evidence of compliance they must be able to provide~\cite{lim2025regulatory}. This regulatory landscape is not only expanding but also becoming more complex. In recent years, the European Union (EU) alone has enacted the General Data Protection Regulation (GDPR)~\cite{gdpr2016}, the AI Act~\cite{euaiact2024}, the Data Act~\cite{dataact2023}, and the Digital Services Act~\cite{dsa2022}, each introducing distinct requirements for software systems. Comparable legislative activity is underway globally, with major jurisdictions introducing their own regulatory frameworks, such as the California Consumer Privacy Act (CCPA/CPRA)~\cite{ccpa2018} in the United States and the Digital Personal Data Protection Act in India~\cite{dpdpa2023}.

Supporting the development of compliant software within this evolving regulatory landscape is widely recognized as a central challenge for requirements engineering (RE) and software engineering, more broadly~\cite{arora2024advancing, abualhaija2024ai}. Despite this recognition, the translation of regulatory provisions into concrete software requirements remains largely manual and error-prone in practice. Empirical evidence highlights the severity of this problem: a recent survey indicates that 90\% of data access requests submitted to EU companies are not fully answered within legally mandated timelines~\cite{noyb2024}. While compliance failures have many causes, one contributing factor is the difficulty of extracting a complete and correct set of software requirements from regulatory text~\cite{breaux2008analyzing}. 

Regulations are written for legal audiences, not for requirements engineers. They are structured around legal concepts, organized to serve legislative and interpretive purposes, and expressed in language that requires substantial interpretation before any engineering activity can begin~\cite{sleimi2021automated}. A requirements engineer seeking to ensure regulatory compliance must navigate a dense mixture of clauses with varying normative character, including obligations, permissions, definitions, recitals, scope declarations, and cross-references. Critically, there is no systematic means for determining which clauses imply software requirements and which do not. This makes regulatory requirements derivation a labor-intensive, error-prone process, and one that must be repeated whenever regulations are amended or new ones are enacted.

To illustrate the challenge, consider Article 17(1) of GDPR~\cite{gdpr2016}: \textit{``The data subject shall have the right to obtain from the controller the erasure of personal data concerning him or her without undue delay.''} Before acting on this provision, a requirements engineer must first recognize that it gives rise to a software requirement. The requirements engineer must then derive one or more software requirements, such as a functional requirement (\textit{``e.g., the system shall delete personal data associated with a data subject upon an approved erasure request''}), a non-functional requirement capturing the timeliness constraint (\textit{``without undue delay''}), and an explanation that allows requirements engineers, developers, testers, and auditors without legal training to understand and trace the requirement back to the source clause. Now consider that the GDPR alone contains 99 articles, and that a requirements engineer may have to look into multiple regulations that are pertinent to a given software project, simultaneously.

Addressing this challenge requires systematic and scalable support for two distinct tasks. The first is identifying which clauses in a regulatory document imply required software system behavior. The second is deriving software requirements from those identified clauses, together with plain-language explanations that justify the derivation and support engineering and audit workflows. These tasks are frequently conflated, yet each presents substantial technical difficulty. Their combination, what we term \textit{regulation-to-requirements translation}, has not, to the best of our knowledge, been addressed as an end-to-end automated pipeline in prior work.


In this paper, we present an automated approach for regulation-to-requirements translation and operationalize it in a supporting tool -- Reg2Req. Our approach identifies requirement-bearing clauses across an entire regulatory document, and then derives software requirements from those clauses, accompanied by explanations that allow requirements engineers, developers, testers, and auditors without legal training to understand and trace each requirement back to its regulatory source. Reg2Req additionally provides cross-reference identification and typing, definition indexing, traceability matrix construction, functional requirement (FR) and non-functional requirement (NFR) classification, and use case seeding to support end-to-end compliance engineering workflows. These features are intended as engineering support for practitioners; they are not core research contributions of this paper and therefore not subject to the same level of empirical evaluation as the primary contributions. A user study with 25 software engineering and requirements engineering practitioners evaluates the impact of the plain-language explanations on requirement comprehension and the perceived utility of Reg2Req's supporting features.

The contributions of this paper are as follows:
\begin{enumerate}
    \item \textbf{Requirement-bearing clause identification.} An automated approach for identifying which clauses in regulatory text imply software requirements.
    
    \item \textbf{Requirement derivation with explanation generation.} An approach for deriving software requirements from identified clauses, accompanied by explanations for requirements engineers, developers, testers, and auditors without legal training.
    
    \item \textbf{Publicly released tool and dataset.} The release of Reg2Req as an open tool, together with annotated datasets for the GDPR and the EU AI Act, in which requirement-bearing clauses have been identified, providing a reusable benchmark for future research.
\end{enumerate}

We examine the following research questions:
\begin{itemize}
    \item \textbf{RQ1:} How accurately does the pipeline identify requirement-bearing clauses?
    \item \textbf{RQ2:} How correct, complete, and clear are the derived requirements and explanations?
    \item \textbf{RQ3:} How does upstream requirement-bearing clause identification affect the quality of derived requirements?
    \item \textbf{RQ4:} To what extent do plain-language explanations support practitioner comprehension of derived requirements?
    \item \textbf{RQ5:} How do practitioners perceive the utility of Reg2Req's supporting features for compliance-related software engineering activities?
\end{itemize}

We evaluate RQ1 and RQ2 on the full clause sets of two regulations from different domains: the GDPR (data protection) and the EU AI Act (AI governance), covering 398 and 574 clauses, respectively. For RQ1, our approach outperforms a SetFit sentence classifier baseline, achieving macro-averaged F1 scores of 0.82 and 0.78. For RQ2, human evaluation shows that derived requirements score highly on completeness across both regulations, while the generated explanations approach ceiling performance on clarity. RQ3 is addressed through an ablation study confirming that the upstream requirement-bearing clause identification stage is a necessary pipeline component. RQ4 and RQ5 are addressed through a user study, which shows that plain-language explanations significantly improve comprehension and confidence in acting on derived requirements, and that Reg2Req's supporting features are perceived as useful for compliance-related software engineering work. We publicly release Reg2Req along with all supporting artifacts, including prompts, datasets, and evaluation material, in a replication package\footnote{\url{https://zenodo.org/records/19209666}}.


\section{Related Work}
\label{sec2}
\subsection{Identifying Requirement-Bearing Text in Regulatory Documents}
A foundational challenge in RE for regulatory compliance is determining which portions of a regulation impose legal obligations, a task referred to in the literature as obligation identification. Breaux and Anton~\cite{breaux2008analyzing} addressed this through semantic parameterization, annotating regulatory text with rights, obligations, and constraints using a structured grammar, but the approach was largely manual and analyst-driven. Kiyavitskaya et al.~\cite{kiyavitskaya2008automating} pursued semi-automation using pattern-based semantic annotation, but showed that obligation-bearing language is frequently embedded in conditional or exception-qualified structures that resist shallow pattern matching~\cite{zeni2015gaiust}.

Deontic logic-based approaches have treated obligation identification more formally. Governatori et al.~\cite{governatori2006compliance} used defeasible logic to capture permission, obligation, and prohibition from legal texts, while Palmirani et al.~\cite{palmirani2018pronto} applied the Akoma Ntoso schema to structure GDPR clauses into machine-readable normative elements. Both approaches require substantial manual curation and do not scale readily to full regulatory documents. Sleimi et al.~\cite{sleimi2021automated} proposed a more automated path using dependency parsing and NLP-based extraction rules for GDPR, but reported persistent difficulties with complex syntactic structures and co-reference. 

Obligation identification, as addressed by the above approaches, is a legal analysis task: it determines whether a clause imposes a normative condition under law. Requirement-bearing clause identification, as defined in this paper, is a RE task: it determines whether a clause specifies, grants, or constrains an observable behavior, interaction, or system property that a software system must implement in order to be compliant. The two tasks are related but distinct. A clause may impose a legal obligation on a human actor or organisation without implying any software system behavior, and conversely, a clause may bear software requirements without containing explicit deontic language. To the best of our knowledge, no prior work frames requirement-bearing clause identification as an explicit, independently evaluable pipeline stage.

\subsection{Requirement Derivation and Explanation Generation from Regulatory Text}
Early work on deriving system requirements from regulatory text was largely manual. Breaux and Anton~\cite{breaux2008analyzing} used their semantic parameterization framework to transform identified obligations into requirement-like statements at significant analyst cost. Goal-oriented approaches offered a structured alternative: NOMOS~\cite{siena2008metamodel, ingolfo2013nomos2} mapped legal permissions and obligations to stakeholder goals and system responsibilities, and Legal GRL~\cite{ghanavati2014legalgrl} introduced regulatory constructs directly into the Goal-oriented Requirement Language. Both frameworks are tightly coupled to specific modeling formalisms, which limits accessibility and resists automation. Zeni et al.~\cite{zeni2015gaiust} partially automated requirement derivation through GaiusT, though it required significant reconfiguration across regulatory regimes.

LLM-based approaches have more recently shown promise. Ioannidis et al.~\cite{ioannidis2023gracenote} used GPT-4 to generate obligation lists from legislative material, and Hassani et al.~\cite{hassani2024rethinking} identified LLMs as a promising direction for automating legal compliance workflows. The most directly relevant work is XTRAREG by Abualhaija et al.~\cite{abualhaija2025xtrareg}, which combines LLMs, prompt engineering, and retrieval-augmented generation to extract privacy requirements from the GDPR for the rights of access and portability, achieving correctness rates of 81.8\% and 85.7\% against 108 expert-extracted reference requirements.

Our approach differs from XTRAREG in three key respects. First, it introduces requirement-bearing clause identification as an explicit upstream stage, operating over an entire regulatory document rather than a pre-selected set of articles. Second, where XTRAREG generates legally grounded rationales targeting legal validation, our approach produces plain-language explanations targeting engineering accessibility for requirements engineers, developers, testers, and auditors without legal training. Third, our approach requires no regulation-specific configuration, whereas XTRAREG is scoped to Articles 15 and 20 of GDPR.

\section{Approach}
\label{sec3}

\begin{figure}[htbp]
\centering
\includegraphics[width=\linewidth]{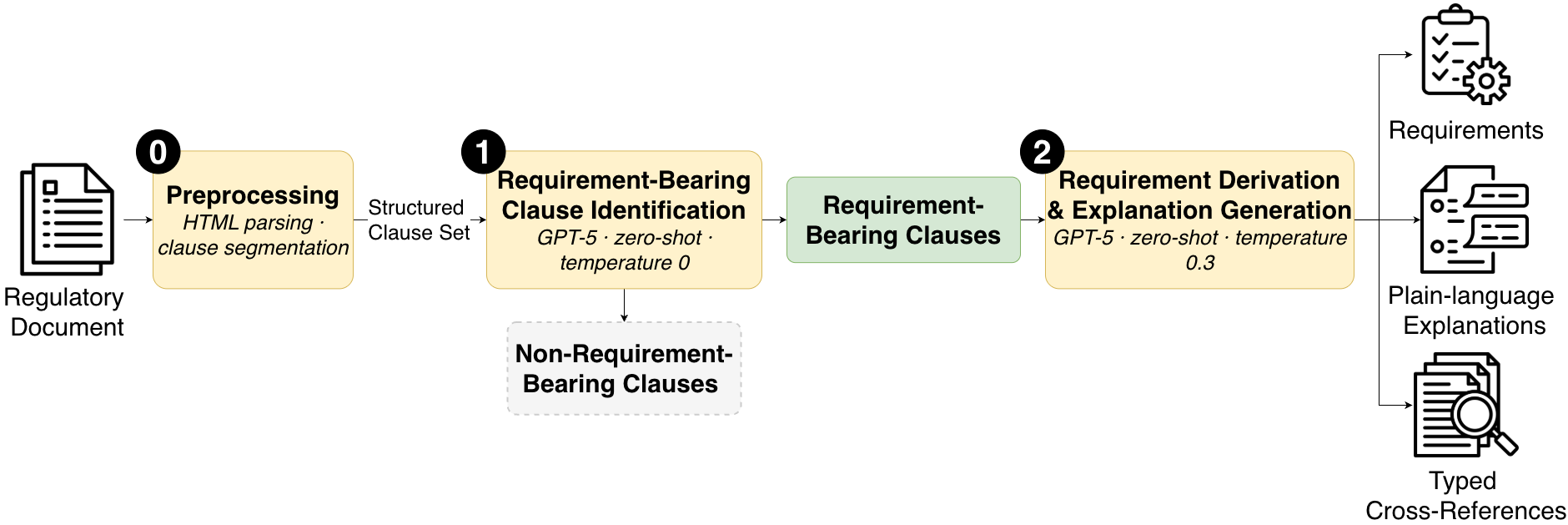}
\caption{Overview of the proposed pipeline.}
\label{fig:pipeline}
\end{figure}
Figure~\ref{fig:pipeline} provides an overview of our approach, which comprises a preprocessing step followed by two sequential stages. In the preprocessing step, a regulatory document is parsed into a structured, clause-level representation. The first stage, requirement-bearing clause identification, filters this clause set to those that imply required system-level behavior for regulatory compliance. Here, system refers to the software-based or software-supported system whose design, implementation, or operation must comply with the regulatory document being processed. The second stage, requirement derivation and explanation generation, transforms each identified clause into one or more system-agnostic (i.e., not tied to any specific technology, architecture, or implementation context) software requirements, accompanied by plain-language explanations that clarify the regulatory intent behind each requirement and support traceability to the originating clause. Next, we explain each of the stages in our approach in detail.


\subsection{Regulatory Document Preprocessing}
The preprocessing stage converts a regulatory document into a structured, clause-level representation that serves as the input to subsequent pipeline stages. It performs two functions: (i) segmenting the document into clauses, and (ii) excluding non-operative content. Clause segmentation is performed at the article-paragraph level. Lettered sub-points (e.g., (a), (b), (c)) are appended to the text of their parent numbered paragraph rather than treated as standalone clauses, as such sub-points typically enumerate conditions, exceptions, or refinements that are governed by the parent provision and are not interpretable in isolation. We exclude recitals in this stage. Under EU drafting conventions, recitals primarily provide interpretive context and legislative intent and generally do not create binding obligations; consequently, they are less suitable as direct sources of system-level requirements~\cite{klimas2008recitals}.


\subsection{Requirement-Bearing Clause Identification}

The first stage of the pipeline determines, for each extracted clause, whether it implies one or more system-level software requirements. A clause is classified as requirement-bearing if it specifies, grants, or constrains an observable behavior, interaction, or system property that a software system must implement in order to be compliant. We formulate this task as a binary classification problem and operationalize it using a zero-shot prompt executed by GPT-5~\cite{openai2025gpt5}. GPT-5 was selected for its strong performance on complex language understanding and generation tasks and its availability through our organisation's approved infrastructure. GPT-5 does not expose temperature as a configurable parameter; output consistency is governed by the model's internal configuration. Each clause is processed independently. 

\begin{figure}[htbp]
\centering
\includegraphics[width=0.65\linewidth]{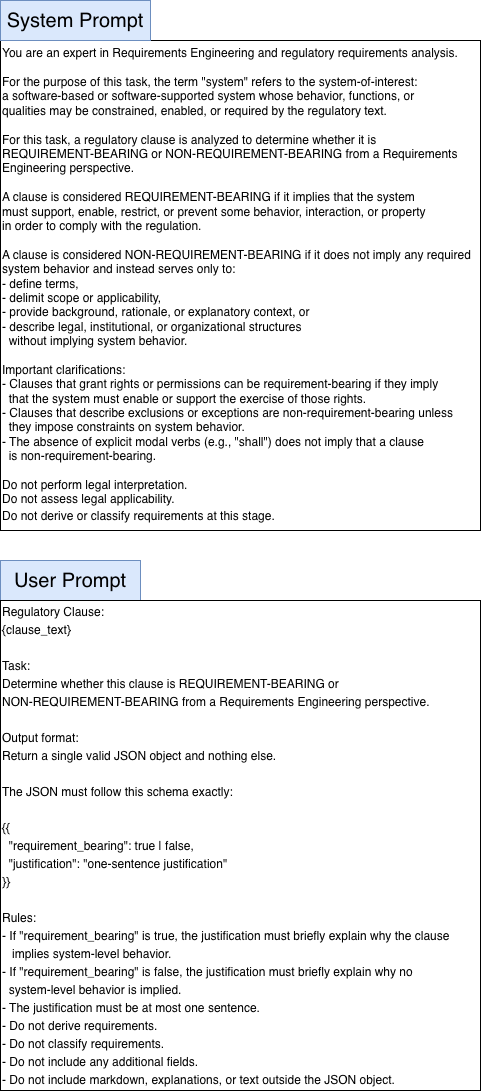}
\caption{Prompt for requirement-bearing clause identification.}
\label{fig:prompt1}
\end{figure}

The prompt, shown in Figure~\ref{fig:prompt1}, was developed through an iterative prompt engineering process~\cite{phoenix2024prompt} and instructs the model to output a binary classification label alongside a one-sentence justification for the classification decision. Requiring an explicit justification was found to improve classification performance, particularly on edge cases such as rights-granting and exception clauses, where prompts requiring only the binary classification label yielded inconsistent results. The final prompt also includes explicit guidance for three recurring edge cases prone to misclassification: rights and permissions clauses, exception clauses, and clauses lacking explicit modal verbs. The model is additionally instructed not to perform legal interpretation. Clauses labeled as requirement-bearing are passed to Stage 2; all others are excluded.

\subsection{Requirement Derivation and Explanation Generation}

The second stage of the pipeline takes each requirement-bearing clause as input and produces three outputs in a single LLM call: derived software requirements, plain-language explanations, and typed cross-reference annotations. These tasks are performed jointly rather than through separate sequential calls for two reasons. First, co-generating requirements and explanations ensures that each explanation is produced with direct awareness of the corresponding requirement text, thereby promoting semantic coherence between the two. Second, cross-reference extraction is a largely surface-level analysis over the clause text; integrating it into the same call adds minimal prompt complexity while avoiding an additional inference pass per clause, thereby reducing cost and latency. This stage also uses GPT-5.


\begin{figure}[htbp]
\centering
\includegraphics[width=0.65\linewidth]{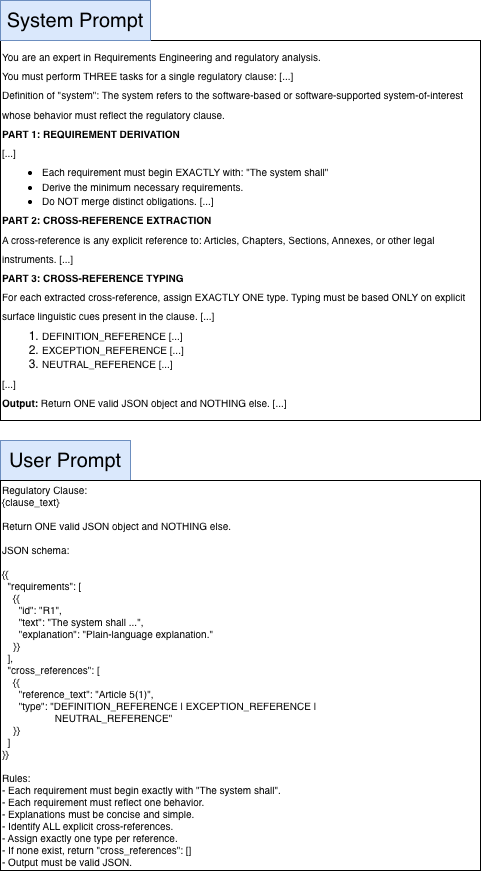}
\caption{Prompt for requirement derivation, explanation generation, and cross-reference extraction and typing.}
\label{fig:prompt2}
\end{figure}

The prompt, illustrated in Figure~\ref{fig:prompt2}, was developed through an iterative prompt engineering process~\cite{phoenix2024prompt} and instructs the model to perform all three tasks: requirement derivation, explanation generation, and cross-reference extraction and typing. For space reasons, the prompt in Figure~\ref{fig:prompt2} is abbreviated; the complete prompt is included in the replication package. For requirement derivation, the model is instructed to derive the minimum necessary set of system-level software requirements implied by the clause, with each requirement describing exactly one concrete system behavior. For explanation generation, the model is instructed to produce a concise, plain-language explanation for each requirement, framed as an engineering-oriented account of what the clause implies for system behavior and why the derived requirement follows from it.  

For cross-reference extraction and typing, the model identifies all explicit references within the clause text to other provisions (e.g., articles, chapters, sections, annexes), or external legal instruments, and assigns each reference a type drawn from a three-class taxonomy. A DEFINITION\_REFERENCE denotes that the clause explicitly imports or relies on a term defined elsewhere; an EXCEPTION\_REFERENCE indicates that the clause limits, qualifies, or conditions its applicability by reference to another provision; and a NEUTRAL\_REFERENCE captures any explicit cross-reference that is structurally informative but does not introduce definitional dependencies or scope-limiting conditions. This taxonomy is a simplified adaptation of cross-reference classification schemes proposed in prior work on legal metadata extraction~\cite{sleimi2021automated}, which distinguish among definitional, conditional, and structural references in legislative text.

We intentionally restrict the taxonomy to three types of cross-references for two reasons. First, finer-grained distinctions generally require legal interpretation to resolve reliably, which is inconsistent with our design decision to rely on surface-level linguistic cues alone. Second, these three types capture the distinctions most relevant for requirement derivation, namely whether a cross-reference imports a definition, limits the scope of applicability, or serves a purely structural role. The typed cross-references are supplementary outputs of the pipeline that are surfaced through the Reg2Req tool to improve traceability.

Both prompts (Figure~\ref{fig:prompt1} and Figure~\ref{fig:prompt2}) were developed through an iterative process involving manual inspection of model outputs on a development set of 30 clauses sampled from the Data Act~\cite{dataact2023}. Successive prompt versions were evaluated by inspecting outputs for recurring failure patterns, which informed the addition of explicit edge case guidance for rights-granting clauses, exception clauses, and clauses lacking modal verbs in the identification prompt, and the instruction to derive the minimum necessary set of requirements in the derivation prompt. No overlap exists between the development set and the evaluation datasets.

\section{Results and Analysis}
\label{sec4}
\subsection{Evaluation Setup}
\subsubsection{Requirement-Bearing Clause Identification}
The evaluation was conducted on the full set of clauses extracted from two EU regulations: the GDPR (398 clauses) and the EU AI Act (574 clauses). We evaluate the pipeline on these acts because both are among the most consequential EU regulations for software systems, with broad compliance obligations across industries, and they differ meaningfully in domain (data protection vs. AI governance) and clause structure, providing a demanding test of regulation-agnostic operation. For the requirement-bearing clause identification stage, ground truth labels were established through manual annotation by six requirements engineers recruited from our organisation, with between 2 and 7 years of professional experience in requirements engineering. None of the annotators had formal legal training. Participation was voluntary, and no financial compensation was provided. Annotation was distributed across annotators, with each clause annotated by a single annotator, and inter-annotator agreement measured on a shared subset of 50 clauses, yielding a Krippendorff's $\alpha$~\cite{krippendorff2004content} of 0.81, indicating strong agreement. Annotation guidelines were provided to ensure consistent labelling criteria across annotators; these are included in the replication package.

We use a SetFit~\cite{tunstall2022setfit} sentence classifier as a baseline for requirement-bearing clause identification because it represents a strong, data-efficient alternative to zero-shot LLM prompting for binary text classification. It has demonstrated competitive performance on short-text classification tasks without requiring large amounts of training data~\cite{tunstall2022setfit}. This makes it a more challenging and realistic baseline than a simple rule-based or keyword-matching approach. The baseline was trained on 80\% of the combined annotated clause set (GDPR and EU AI Act, stratified by label) using the \textit{paraphrase-mpnet-base-v2}~\cite{reimers2019sentencebert} sentence transformer as the base model, and evaluated on the remaining 20\% held-out test split ($n = 195$). The sentence transformer was fine-tuned contrastively with a learning rate of $2 \times 10^{-5}$, batch size 16, 1 epoch, and maximum sequence length 256, using the AdamW optimiser with weight decay 0.01 and a dropout rate of 0.1. The classification head was a logistic regression model trained on the resulting embeddings using default \textit{scikit-learn}~\cite{pedregosa2011scikit} hyperparameters ($C = 1.0$, solver: \texttt{lbfgs}). Our approach was evaluated on the full clause set and on the same 20\% held-out split to enable direct comparison with the SetFit baseline.

\subsubsection{Requirement Derivation and Explanation Generation}
The pipeline produces 448 requirements and explanations from GDPR and 643 from the EU AI Act (1,091 in total). The same six requirements engineers rated all 1,091 derived requirements and their accompanying explanations on three dimensions each (correctness, completeness, and clarity on a 1--5 scale with rubric-anchored criteria). Annotators were provided with detailed annotation guidelines prior to the evaluation; these are included in the replication package.

For requirements, \textit{correctness} captures whether the derived requirement accurately reflects the content of the source clause without introducing unsupported statements, and is evaluated at the requirement level. \textit{Completeness} captures whether all distinct behaviors, interactions, and properties implied by the clause have been captured and is evaluated at the clause level, considering all requirements derived from a given clause collectively. \textit{Clarity} captures whether the derived requirement is unambiguous and comprehensible to an individual without legal training and is evaluated at the requirement level.

For explanations, \textit{correctness} captures whether the explanation accurately reflects the link between the source clause and the derived requirement. \textit{Completeness} captures whether the explanation adequately covers all reasoning linking the source clause to the derived requirement. \textit{Clarity} captures whether the explanation is comprehensible to a requirements engineer, developer, tester, or auditor without legal training. All explanation metrics are evaluated at the requirement level, as each explanation corresponds to a single derived requirement. \textit{Plausibility}, defined as whether an explanation is reasonable and supported by the requirement, was assessed using a binary rating. All explanations were rated as plausible, consistent with the findings of Abualhaija et al.~\cite{abualhaija2025xtrareg} for XTRAREG-generated rationales.

Evaluation load was distributed equally across annotators, with each output rated by a single evaluator. Inter-rater agreement was measured on a shared overlap subset of 30 requirement--explanation pairs across six raters, yielding per-metric Krippendorff's $\alpha$ values of 0.76 for requirement correctness, 0.79 for requirement completeness, 0.81 for requirement clarity, 0.74 for explanation correctness, 0.72 for explanation completeness, and 0.71 for explanation clarity. Plausibility yielded perfect agreement ($\alpha = 1.0$), consistent with the finding that all explanations were rated as plausible.

\subsubsection{Ablation Study}
\label{sec:ablation}
The ablation study compares the full pipeline against a variant in which all clauses are fed directly to the requirement derivation stage without prior requirement-bearing clause identification. In this variant, the derivation prompt was modified to instruct the model to derive requirements only if the clause implies system-level software requirements, returning an empty list otherwise. Since this variant processes all clauses, the resulting output set is substantially larger, making human evaluation infeasible. We therefore use GPT-5.4~\cite{openai2025gpt54} as an automated judge~\cite{zheng2023judging}, with temperature set to 0 for deterministic outputs. To validate this approach, we computed Spearman rank correlations~\cite{spearman1904proof} between GPT-5.4 judge scores and human ratings on the RQ2 outputs, obtaining strong correlations across all metrics ($\rho = 0.77$--$0.85$), confirming that GPT-5.4 judgments are a valid proxy for human evaluation. The LLM-as-a-judge prompt is provided in the replication package. Cross-reference extraction and typing were evaluated using the same setup as a supplementary analysis.

\subsection{RQ1: How accurately does the approach identify requirement-bearing clauses?}

\begin{table}[t]
\centering
\caption{Requirement-bearing clause identification results (macro-averaged)}
\label{tab:full_results}
\scriptsize
\setlength{\tabcolsep}{4pt}
\begin{tabular}{lccccc}
\hline
\textbf{Dataset} & \textbf{Accuracy} & \textbf{Precision} & \textbf{Recall} & \textbf{F1} & \textbf{Support} \\
\hline
GDPR & 0.832 & 0.804 & 0.867 & 0.815 & 398 \\
EU AI Act & 0.803 & 0.772 & 0.871 & 0.779 & 574 \\
\hline
\end{tabular}
\end{table}

\begin{table}[t]
\centering
\caption{Requirement-bearing clause identification results on the held-out test split (macro-averaged)}
\label{tab:heldout_results}
\scriptsize
\setlength{\tabcolsep}{3pt}
\begin{tabular}{lccccc}
\hline
\textbf{Model} & \textbf{Accuracy} & \textbf{Precision} & \textbf{Recall} & \textbf{F1} & \textbf{Support} \\
\hline
Our Approach & 0.821 & 0.793 & 0.854 & 0.822 & 195 \\
SetFit (paraphrase-mpnet-base-v2) & 0.756 & 0.684 & 0.728 & 0.705 & 195 \\
\hline
\end{tabular}
\end{table}

Table~\ref{tab:full_results} reports macro-averaged requirement-bearing clause identification results on the full clause sets of both regulations. The approach achieves an F1 of 0.815 on GDPR and 0.779 on the EU AI Act. Recall is consistently high across both regulations (0.867 and 0.871), indicating that the approach rarely misses a requirement-bearing clause. The dominant error type is false positives (61 for GDPR and 113 for the EU AI Act), reflecting a tendency to classify borderline clauses as requirement-bearing. For example, GDPR-1-3 (``The free movement of personal data within the Union shall neither be restricted nor prohibited...'') was classified as requirement-bearing because it constrains data transfer behavior, but was annotated as non-requirement-bearing since it addresses a legislative principle rather than a system obligation. Table~\ref{tab:heldout_results} reports results on the 20\% held-out test split, comparing our approach against the SetFit-based baseline. Our approach outperforms SetFit on all metrics ($n = 195$), with an F1 gap of 0.117 (0.822 vs.\ 0.705) reflecting consistent improvements across both precision and recall. While false negatives are rare, missed requirement-bearing clauses are silently excluded from the pipeline, which may undermine practitioner trust in the completeness of the derived requirements. Future work could explore confidence scoring or human-in-the-loop review for borderline clauses to make such omissions visible and actionable.


\subsection{RQ2: How correct, complete, and clear are the derived requirements and explanations?}

\begin{table}[t]
\centering
\caption{Requirement derivation and explanation generation: human evaluation results (1--5 scale)}
\label{tab:human_eval}
\scriptsize
\setlength{\tabcolsep}{3pt}
\begin{tabular}{llccccc}
\hline
\textbf{Category} & \textbf{Metric} & \multicolumn{2}{c}{\textbf{GDPR}} & \multicolumn{2}{c}{\textbf{EU AI Act}} & \textbf{Level} \\
 &  & \textbf{Mean} & \textbf{SD} & \textbf{Mean} & \textbf{SD} & \\
\hline
\multirow{3}{*}{Requirements}
& Correctness & 3.74 & 0.87 & 3.54 & 0.81 & per-req \\
& Completeness & 4.60 & 0.76 & 4.45 & 0.85 & per-clause \\
& Clarity & 4.27 & 0.75 & 4.31 & 0.68 & per-req \\
\hline
\multirow{3}{*}{Explanations}
& Correctness & 4.06 & 0.69 & 3.85 & 0.69 & per-req \\
& Completeness & 4.36 & 0.82 & 4.18 & 0.89 & per-req \\
& Clarity & 4.92 & 0.28 & 4.94 & 0.24 & per-req \\
\hline
\end{tabular}
\end{table}

Table~\ref{tab:human_eval} reports human evaluation scores for requirement derivation and explanation generation across both regulations. The \textit{Level} column indicates whether scoring was conducted per-requirement or per-clause. Explanation clarity achieves the highest scores across both regulations (4.92 $\pm$ 0.28 for GDPR, 4.94 $\pm$ 0.24 for the EU AI Act), approaching ceiling with very low standard deviations. Requirement completeness is also high (4.60 $\pm$ 0.76 for GDPR, 4.45 $\pm$ 0.85 for the EU AI Act).

Requirement correctness is the weakest dimension on both regulations (3.74 $\pm$ 0.87 for GDPR, 3.54 $\pm$ 0.81 for the EU AI Act), with the highest standard deviations reflecting greater variability. Lower scores arise primarily in clauses whose implied requirements are qualified by exceptions or conditions, and in rights-granting clauses where the system-level requirements are less directly stated. For instance, GDPR-28-3/R5 received a correctness score of 1 because the derived requirement (``The system shall respect the conditions referred to in paragraphs 2 and 4'') merely repeats an unresolved cross-reference rather than translating the conditions into concrete system behavior. The EU AI Act scores lower than GDPR on both requirement and explanation correctness, consistent with the greater technical complexity of that regulation.

A direct quantitative comparison with XTRAREG~\cite{abualhaija2025xtrareg} is not meaningful for three reasons: (1) XTRAREG evaluates against a reference set compiled from sources beyond the GDPR, whereas our pipeline uses the regulatory clause text as the sole input; (2) XTRAREG's evaluation was conducted by a legal expert using legal correctness criteria, whereas ours uses RE practitioners; and (3) XTRAREG targets two pre-selected GDPR rights, whereas our pipeline operates across full clause sets of regulations, without pre-selection.


\subsection{RQ3: How does upstream requirement-bearing clause identification affect the quality of derived requirements?}

\begin{table}[t]
\centering
\caption{Ablation: full pipeline vs. without identification stage}
\label{tab:ablation}
\scriptsize
\setlength{\tabcolsep}{2pt}
\begin{tabular}{llcccccc}
\hline
\textbf{Category} & \textbf{Metric} & \multicolumn{3}{c}{\textbf{GDPR}} & \multicolumn{3}{c}{\textbf{EU AI Act}} \\
 &  & \textbf{Full} & \textbf{No ident.} & $\Delta$ & \textbf{Full} & \textbf{No ident.} & $\Delta$ \\
\hline
\multirow{3}{*}{Req.}
& Correctness & 3.81$\pm$0.92 & 2.81$\pm$0.94 & -1.00 & 3.61$\pm$0.85 & 2.63$\pm$0.89 & -0.98 \\
& Completeness & 4.65$\pm$0.79 & 3.42$\pm$1.02 & -1.23 & 4.51$\pm$0.88 & 3.28$\pm$0.97 & -1.23 \\
& Clarity & 4.31$\pm$0.78 & 3.89$\pm$0.81 & -0.42 & 4.35$\pm$0.71 & 3.76$\pm$0.79 & -0.59 \\
\hline
\multirow{3}{*}{Expl.}
& Correctness & 4.12$\pm$0.73 & 3.21$\pm$0.88 & -0.91 & 3.91$\pm$0.72 & 3.02$\pm$0.84 & -0.89 \\
& Completeness & 4.29$\pm$0.86 & 3.18$\pm$1.05 & -1.11 & 4.11$\pm$0.93 & 3.05$\pm$0.98 & -1.06 \\
& Clarity & 4.89$\pm$0.31 & 4.54$\pm$0.61 & -0.35 & 4.91$\pm$0.27 & 4.49$\pm$0.67 & -0.42 \\
\hline
\end{tabular}
\end{table}

Table~\ref{tab:ablation} reports LLM-as-a-judge scores (GPT-5.4) comparing the full pipeline against the ablated variant in which all clauses are fed directly to the requirement derivation stage without prior requirement-bearing clause identification. Full pipeline scores are reported as LLM judge scores rather than human scores to ensure a like-for-like comparison; these differ slightly from the human evaluation scores in Table~\ref{tab:human_eval}, consistent with the Spearman rank correlations reported in Section~\ref{sec:ablation} ($\rho = 0.77$--$0.85$ across metrics). The full pipeline LLM judge scores are marginally higher than human scores on most metrics, reflecting a slight positive bias typical of LLM judges~\cite{zheng2023judging}.

Removing the requirement-bearing clause identification stage produces substantial drops across all dimensions on both regulations. The largest drops are in requirement completeness (GDPR: $-1.23$, EU AI Act: $-1.23$) and explanation completeness (GDPR: $-1.11$, EU AI Act: $-1.06$), with notably higher standard deviations in the ablated condition indicating greater output variability. Clarity shows the smallest drops, confirming that the requirement derivation stage produces clear outputs regardless of input quality. These results confirm that the upstream requirement-bearing clause identification stage is a necessary pipeline component.

\subsection{Supplementary Analysis: Cross-reference Extraction and Typing}
Table~\ref{tab:xref} reports cross-reference extraction and typing results. Extraction precision is near-perfect on both regulations (99.49\% for GDPR, 100.00\% for the EU AI Act), while recall is lower (84.48\% and 87.43\%), with missing references primarily attributable to cross-references embedded in non-standard syntactic structures. For example, the reference to `paragraphs 1 and 2' in GDPR-14-3 was missed, as it appears as an internal paragraph reference rather than an explicit article citation. Typing accuracy is high on both regulations (92.89\% and 95.42\%), confirming that the three-type taxonomy is reliably applicable from surface linguistic cues alone.
\begin{table}[h]
\centering
\caption{Cross-reference extraction and typing results}
\label{tab:xref}
\scriptsize
\setlength{\tabcolsep}{4pt}
\begin{tabular}{lcc}
\hline
\textbf{Metric} & \textbf{GDPR} & \textbf{EU AI Act} \\
\hline
Extracted references & 197 & 306 \\
Correctly extracted & 196 & 306 \\
Missing references (FN) & 36 & 44 \\
Extraction Precision & 99.49\% & 100.00\% \\
Extraction Recall & 84.48\% & 87.43\% \\
Extraction F1 & 91.38\% & 93.29\% \\
Typing Accuracy & 92.89\% (183/197) & 95.42\% (292/306) \\
\hline
\end{tabular}
\end{table}

\section{Reg2Req}
\label{sec5}
Reg2Req is an interactive tool that operationalizes the two-stage pipeline described in Section~\ref{sec3} within a compliance engineering workflow. A requirements engineer can load a regulatory document, execute the pipeline, and explore the resulting artefacts through a structured user interface. The tool provides direct access to the core pipeline outputs: identified requirement-bearing clauses, derived system-level software requirements, and their accompanying plain-language explanations alongside typed cross-references extracted as a supplementary pipeline output.  

In addition to these, Reg2Req offers four supporting features designed to assist end-to-end compliance engineering: functional and non-functional requirement (FR/NFR) classification, use case seeding, traceability matrix navigation, and definition indexing. These capabilities are intended to facilitate downstream engineering, analysis, and audit activities by enhancing organization, navigation, and interpretability of the derived requirements. While these features contribute to practitioner usability and workflow integration, they are considered engineering support features rather than primary research contributions and are therefore not subject to the same level of empirical evaluation as the two main pipeline stages.

\subsection{FR/NFR Classification}

\begin{figure}[htbp]
\centering
\includegraphics[width=\linewidth]{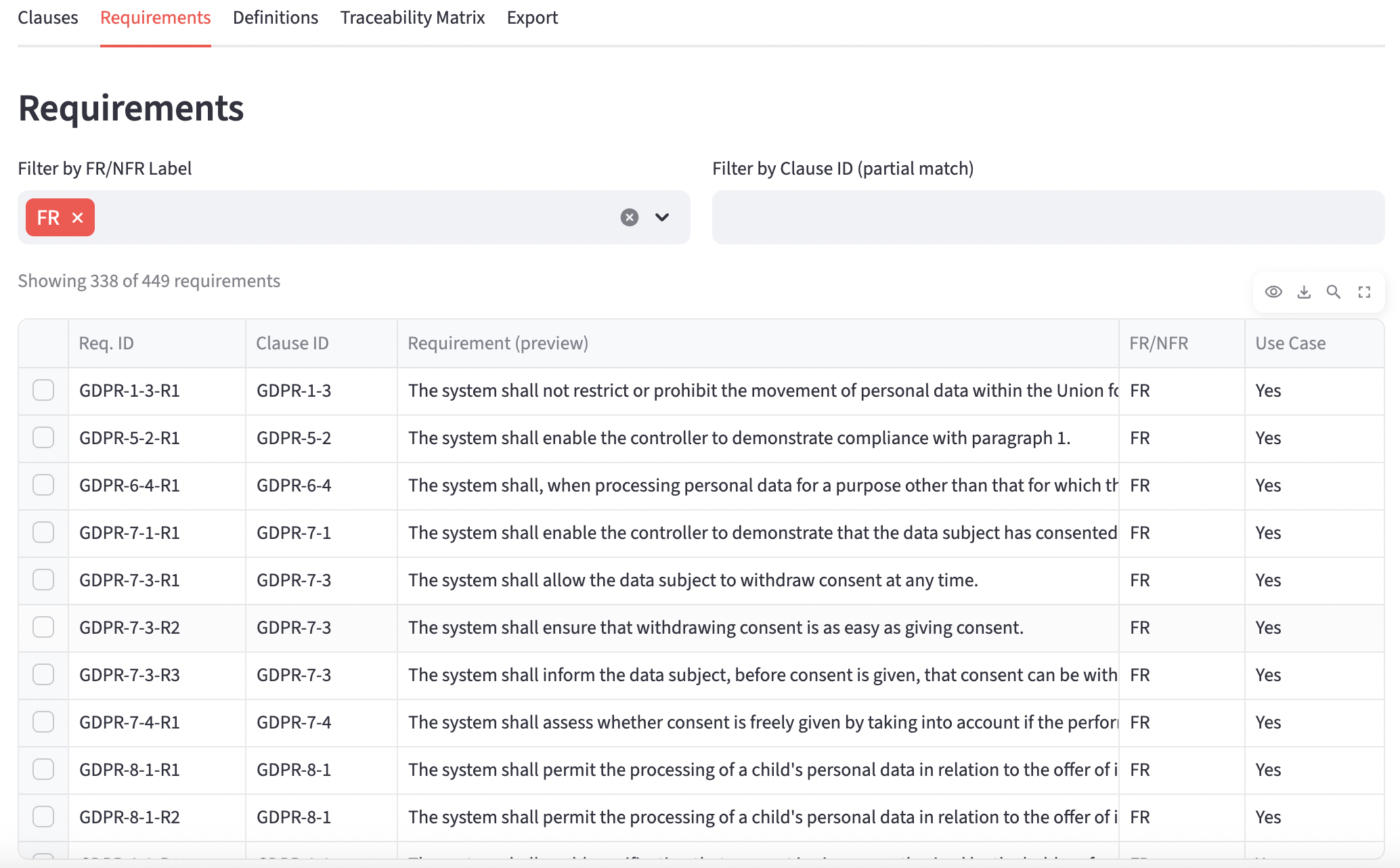}
\caption{The Requirements tab in Reg2Req, filtered to show FR-labelled requirements.}
\label{fig:FR}
\end{figure}

Each derived requirement is automatically classified as functional (FR), non-functional (NFR), or both (FR+NFR) using GPT-5~\cite{openai2025gpt5}. Ground truth labels were established through manual annotation of all derived requirements across both regulations by the same six evaluators who participated in the human evaluation of derived requirements and generated explanations. The evaluators applied the FR/NFR classification scheme adapted from Dalpiaz et al.~\cite{dalpiaz2019requirements}. Inter-annotator agreement was assessed on an overlapping subset of 50 requirements, yielding a Krippendorff's $\alpha$ of 0.72, indicating substantial agreement. Per-label and macro-averaged precision, recall, and F1 scores against the annotated ground truth are reported in Table~\ref{tab:frnfr}.
\begin{table}[h]
\centering
\caption{FR/NFR Classification Performance}
\label{tab:frnfr}
\setlength{\tabcolsep}{5pt}
\begin{tabular}{llcccc}
\hline
\textbf{Dataset} & \textbf{Label} & \textbf{P} & \textbf{R} & \textbf{F1} & \textbf{Support} \\
\hline
\multirow{5}{*}{GDPR} 
& FR & 0.914 & 0.871 & 0.892 & 356 \\
& NFR & 0.750 & 0.180 & 0.290 & 50 \\
& FR+NFR & 0.351 & 0.810 & 0.489 & 42 \\
& Macro & 0.672 & 0.620 & 0.557 & -- \\
& Weighted & 0.843 & 0.788 & 0.787 & -- \\
\hline
\multirow{5}{*}{EU AI Act} 
& FR & 0.624 & 0.961 & 0.757 & 308 \\
& NFR & 0.692 & 0.409 & 0.514 & 67 \\
& FR+NFR & 0.815 & 0.396 & 0.533 & 268 \\
& Macro & 0.711 & 0.589 & 0.601 & -- \\
& Weighted & 0.711 & 0.668 & 0.638 & -- \\
\hline
\end{tabular}
\end{table}
Classification performance is stronger for GDPR (overall accuracy 78.8\%) than for the EU AI Act (66.7\%). Across both regulations, functional requirements are classified most reliably, reflecting their syntactic distinctiveness. The NFR and FR+NFR classes prove harder to separate, particularly on GDPR, where NFR recall is low (0.18), likely because regulatory text tends to embed quality constraints implicitly within functional obligations rather than expressing them as standalone non-functional statements. Misclassification of NFR as FR may result in use case seeds being generated for requirements that are better addressed through architectural or quality decisions, potentially misleading downstream engineering activities. The FR/NFR classification is intended as a lightweight decision-support feature rather than a definitive classification; practitioners are encouraged to review classifications before acting on them. Figure~\ref{fig:FR} shows the Requirements tab in Reg2Req, illustrating how practitioners can navigate derived requirements by classification type.

\subsection{Use Case Seeding}
For each requirement classified as FR or FR+NFR, Reg2Req generates a partially populated use case seed using GPT-5 with a few-shot prompt. Each seed follows a fixed schema adapted from Cockburn et al.~\cite{cockburn2001usecases}, comprising a use case name, a goal statement, and a list of system responsibilities. Actors, interaction flows, preconditions, and postconditions are intentionally omitted, as the seeds are designed to provide a compliance-grounded starting point for further elicitation rather than fully specified use cases. The prompt prohibits the model from assuming specific technologies, actors, interfaces, or data models, keeping the output implementation-neutral.

\subsection{Traceability Matrix}
\begin{figure}[htbp]
\centering
\includegraphics[width=\linewidth]{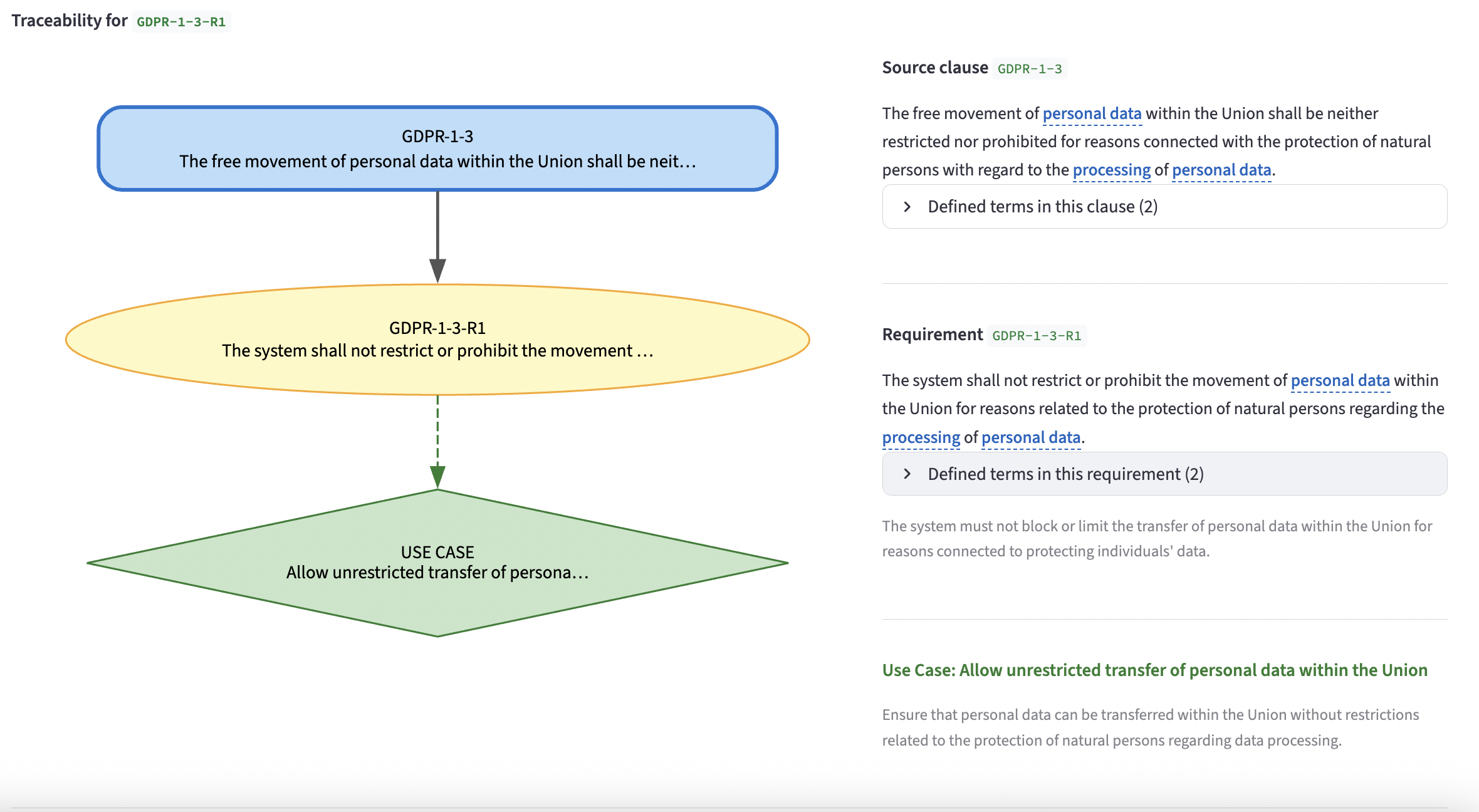}
\caption{Traceability view for requirement GDPR-1-3-R1 in Reg2Req. The left panel shows the traceability chain from the source clause to the derived requirement to use case seed. The right panel displays the full clause text with hyperlinked defined terms, the derived requirement with its plain-language explanation, and the use case seed name and goal statement.}
\label{fig:traceability}
\end{figure}

Reg2Req provides end-to-end traceability between regulatory clauses and derived artifacts through a two-view traceability matrix. In the requirement view, a practitioner can view all derived requirements. Upon selecting a derived requirement the practitioner is presented with its originating clause, associated cross-references together with their types, its FR/NFR classification, and, where applicable, the corresponding use case seed. In the clause view, a practitioner can view all regulatory clauses. When the practitioner selects a regulatory clause, they can observe all downstream requirements derived from it, along with their associated use case seeds when available, and cross-references together with their types. Figure~\ref{fig:traceability} illustrates the traceability view for a requirement derived from GDPR Article 1, Paragraph 3.

\begin{figure}[h]
\centering
\includegraphics[width=\linewidth]{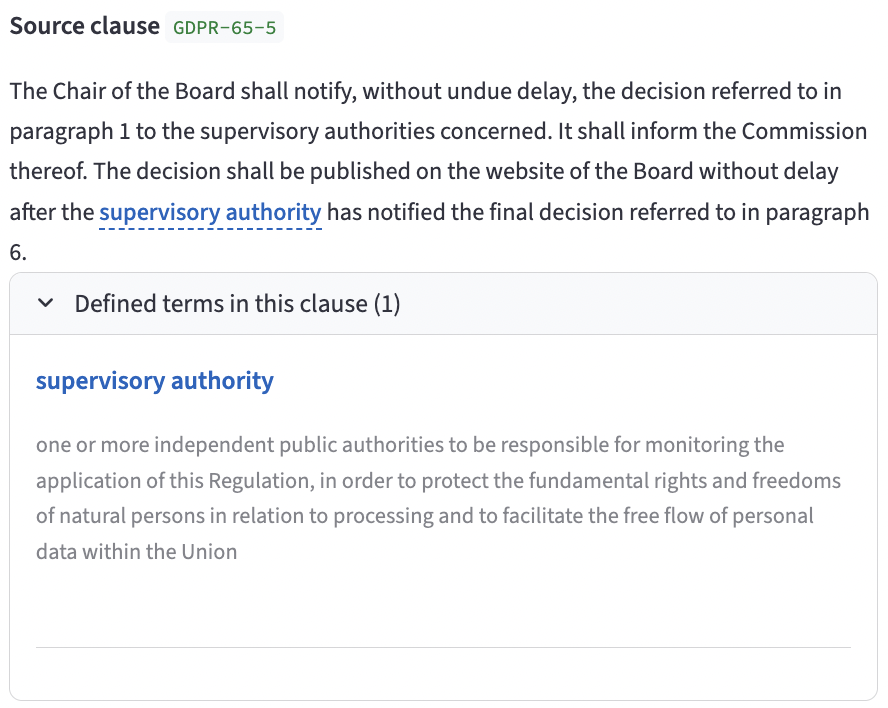}
\caption{Definition indexing in Reg2Req. Defined terms appearing in a clause are hyperlinked inline, and their definitions are surfaced in an expandable panel below the clause text.}
\label{fig:definition}
\end{figure}
\subsection{Definition Indexing}
Reg2Req extracts and indexes explicit term definitions from the regulatory document using GPT-5. Definition extraction is performed over all clauses, rather than being limited to designated definition articles. The model is instructed to extract definitions verbatim only when the text explicitly defines a term using clear surface cues such as ``means'', ``is defined as'', or ``refers to''. The output is a structured glossary consisting of (term, definition, source clause) triples. In the Reg2Req user interface, terms for which definitions have been extracted are highlighted throughout the tool. Selecting a highlighted term reveals its definition along with the originating clause. Where a term is defined multiple times — as occurs in regulations that redefine concepts for specific contexts — all definitions are displayed together. Figure~\ref{fig:definition} illustrates the definition indexing feature.

\section{User Study}
\label{sec6}
We conducted an online questionnaire-based survey study to evaluate the extent to which plain-language explanations support practitioner comprehension of derived requirements (RQ4), and how practitioners perceive the utility of Reg2Req’s supporting features for compliance-related software engineering activities (RQ5). We recruited 25 software engineers and requirements engineers from professional networks and academic contacts. Participation was voluntary, and no financial compensation was provided. No legal expertise was required or expected, reflecting the target user population of Reg2Req. 

The survey comprised five sections and was administered online via Google Forms. Section 1 collected demographic information, including role, years of professional experience, familiarity with regulations and regulatory requirements, and prior experience with compliance tooling. Sections 2 and 3 formed a within-participants before/after design. Participants were presented with five requirements derived by Reg2Req from our evaluation dataset, selected to represent a range of complexity and clause types across both GDPR and the EU AI Act. In Section 2, participants rated each requirement on comprehension (1–5) and confidence in acting on the requirement (1–5), and indicated whether they could identify a concern or ambiguity. In Section 3, the same requirements were shown again with their plain-language explanations generated by Reg2Req, and participants repeated the same questions; they additionally indicated whether the explanation revealed a concern not noticed in Section 2. Section 4 presented participants with an example output screenshot for each of five Reg2Req supporting features: FR/NFR classification of derived requirements, use case seed generation, cross-reference identification and typing, definition indexing, and the traceability matrix. For each feature, participants rated perceived usefulness on a 1–5 Likert scale and indicated whether they would use the feature when working with regulations in a software project. Section 5 collected an overall assessment, including a question on whether they would use Reg2Req as a starting point for deriving software requirements from a new regulation, and two open-ended questions inviting general feedback and suggested improvements. The complete questionnaire and anonymized survey responses are provided in our replication package.

\subsection{Impact of Explanations on Practitioner Comprehension}
Study participants were predominantly software developers (15, 60\%) and requirements engineers or analysts (8, 32\%). Experience ranged across all levels, with the majority having between 2 and 10 years of professional experience (18, 72\%). Most participants had some exposure to regulations in their work; 22 (88\%) reported working with regulations at least occasionally. Nine participants (36\%) had prior experience with compliance tooling, of whom 4 indicated their prior tool provided explanations alongside its outputs.

Comprehension of the derived requirements increased from 3.29 (SD = 0.66) to 4.17 (SD = 0.55) after explanation exposure ($\Delta = +0.88$, $p < 0.001$), indicating that participants found the requirements substantially clearer when accompanied by plain-language explanations. Confidence in acting on the requirements, whether implementing, testing, or specifying them further, increased from 2.44 (SD = 0.77) to 3.42 (SD = 0.73) ($\Delta = +0.98$, $p < 0.001$), the largest improvement observed across all measures. The concern identification rate increased from 40.0\% to 59.2\%, suggesting that explanations enabled participants to detect implementation ambiguities and edge cases that were not apparent from the requirement text alone. All significance values are from the Wilcoxon signed-rank test\cite{wilcoxon1945individual} applied to the paired before/after scores. 

Qualitative themes from open-ended responses included the clarity of legal intent conveyed by explanations (participants noted that explanations clarified ambiguous regulatory terms such as `undue delay' and `proportionate') and a reduced need to consult the original regulation when acting on derived requirements.

\subsection{Perceived Utility of Reg2Req Supporting Features}

\begin{table}[t]
\centering
\caption{Perceived usefulness (1--5) and intended adoption of Reg2Req supporting features}
\label{tab:usefulness}
\scriptsize
\setlength{\tabcolsep}{2pt}
\begin{tabular}{lcc}
\hline
\textbf{Feature} & \textbf{Usefulness (mean $\pm$ SD)} & \textbf{Would use (\%)} \\
\hline
FR/NFR classification & 4.16 $\pm$ 0.69 & 100 \\
Use case seed generation & 3.96 $\pm$ 0.79 & 96 \\
Cross-reference identification and typing & 3.48 $\pm$ 0.92 & 92 \\
Definition indexing & 4.40 $\pm$ 0.58 & 100 \\
Traceability matrix & 4.64 $\pm$ 0.57 & 100 \\
\hline
\end{tabular}
\end{table}

Table~\ref{tab:usefulness} reports mean perceived usefulness ratings and the proportion of participants who indicated they would use each feature to support compliance-related work. `Would use' reports the combined percentage of participants who responded `Yes' or `Possibly' to the question of whether they would use the feature when working with regulations in a software project.

The traceability matrix feature received the highest perceived usefulness rating (4.64 $\pm$ 0.57), with all 25 participants (100\%) indicating they would use it when working with regulatory requirements. Definition indexing was rated second highest (4.40 $\pm$ 0.58), also with 100\% indicating they would use it. The cross-reference identification and typing feature received the lowest rating (3.48 $\pm$ 0.92); qualitative feedback suggested that participants were less certain how to incorporate cross-reference type distinctions into existing engineering workflows, though 92\% still indicated they would use it. Developers tended to rate the traceability matrix and definition indexing most highly, while requirements engineers additionally placed strong value on the FR/NFR classification and cross-reference features.

When asked whether they would use Reg2Req as a starting point for deriving software requirements from a new regulation, all 25 participants responded yes (80\%) or possibly (20\%). Open-ended feedback highlighted a strong desire for integration with existing requirements management and software development tools, and a strong interest in extending support to additional regulations, reflecting practitioners' intent to use Reg2Req beyond the evaluated scope.

\section{Threats to Validity}
\label{sec7}

\subsection{Construct Validity}
A central construct validity threat concerns the operationalization of requirement-bearing clause identification. Our operationalization represents one reasonable interpretation, but alternatives are possible for edge cases such as rights-granting and exception clauses; Krippendorff's $\alpha$ = 0.81 confirms consistent application of the guidelines. A second threat concerns the evaluation metrics: correctness, completeness, and clarity lack standardized definitions in RE or legal NLP, mitigated through explicit metric definitions; per-metric $\alpha$ values of 0.71--0.81 confirm consistent application. The pipeline derives requirements from clause text alone without resolving cross-references, which may affect quality for clauses with heavy cross-referenced content. Finally, RQ5 feature utility was assessed via screenshots rather than live tool interaction, which may not fully capture real-world perceptions; future work should conduct task-based studies with the live tool.



\subsection{Internal Validity}
The primary internal validity threat concerns annotation quality. Clauses were distributed across annotators with a single annotator per clause, mitigated by explicit annotation guidelines and Krippendorff's $\alpha$ = 0.81 on the shared subset. As annotators were requirements engineers without legal training, ground truth labels represent an RE interpretation of regulatory text rather than a legally authoritative one; disagreements resolvable only by a legal expert may be conflated with annotation ambiguity. For the human evaluation of derived requirements and explanations, none of the annotators had prior experience with GDPR or the EU AI Act, reducing the risk of domain familiarity biasing ratings; per-metric $\alpha$ values of 0.71--0.81 confirm consistent rating behaviour.

\subsection{External Validity}

The evaluation is conducted on two EU regulations — GDPR and the EU AI Act — which, despite differing in domain, share structural characteristics arising from EU legislative drafting conventions; generalisation to other jurisdictions or legal traditions cannot be assumed. Both regulations also evolve relatively slowly, limiting the immediate benefit of re-execution for regulatory change, though automation benefits are most pronounced when organisations must track a large and expanding set of regulations simultaneously. A further threat concerns the use of GPT-5 and GPT-5.4, selected for their strong performance and availability through our organisation's approved infrastructure; generalisation to other models, including open-source LLMs, remains an open question. LLM behaviour is known to drift over time~\cite{chen2024chatgpt}, potentially affecting reproducibility. We mitigate both threats by publicly releasing all prompts and annotated datasets, enabling replication with alternative models or future model versions.
\subsection{Conclusion Validity}
The ablation study uses LLM-as-a-judge scores as a proxy for human evaluation to assess the effect of upstream requirement-bearing clause identification on the quality of derived requirements. The validity of conclusions drawn from the ablation study depends on the degree to which LLM judge scores correlate with human ratings. We validated this by computing Spearman $\rho$ between LLM-as-a-judge scores and human ratings on the same set of requirements derived by the full pipeline, obtaining values ranging from 0.77 to 0.85 across all evaluation metrics. This provides reasonable confidence that the LLM judge captures the same quality dimensions as human raters. 


\section{Conclusion}
\label{sec8}
This paper presents Reg2Req, an automated pipeline for translating regulatory text into system-level software requirements, evaluated on the full clause sets of GDPR and the EU AI Act. Requirement-bearing clause identification achieves macro-averaged F1 scores of 0.82 and 0.78, outperforming a SetFit baseline, and human evaluation confirms high completeness and near-ceiling explanation clarity. A user study with 25 practitioners shows that plain-language explanations significantly improve comprehension and confidence, and that Reg2Req's supporting features are perceived as useful for compliance-related work. Beyond initial derivation, the pipeline supports ongoing compliance activities: when a regulation is amended, Reg2Req can be re-executed on the updated clause set, with traceability links enabling engineers to identify affected requirements. The approach requires no regulation-specific configuration, making it applicable to newly enacted regulations and suitable for organisations tracking an expanding regulatory landscape. Future work includes retrieval-augmented generation to improve correctness in clauses with unresolved cross-references, and evaluation on regulations beyond EU law. 

\bibliographystyle{IEEEtran}
\bibliography{references}

@incollection{arora2024advancing,
  title={Advancing requirements engineering through generative ai: Assessing the role of llms},
  author={Arora, Chetan and Grundy, John and Abdelrazek, Mohamed},
  booktitle={Generative AI for Effective Software Development},
  pages={129--148},
  year={2024},
  publisher={Springer}
}

@inproceedings{abualhaija2024ai,
  title={AI-enabled regulatory change analysis of legal requirements},
  author={Abualhaija, Sallam and Ceci, Marcello and Sannier, Nicolas and Bianculli, Domenico and Briand, Lionel C and Zetzsche, Dirk and Bodellini, Marco},
  booktitle={2024 IEEE 32nd International Requirements Engineering Conference (RE)},
  pages={5--17},
  year={2024},
  organization={IEEE}
}

@article{kosenkov2025systematic,
  title={Systematic mapping study on requirements engineering for regulatory compliance of software systems},
  author={Kosenkov, Oleksandr and Elahidoost, Parisa and Gorschek, Tony and Fischbach, Jannik and Mendez, Daniel and Unterkalmsteiner, Michael and Fucci, Davide and Mohanani, Rahul},
  journal={Information and Software Technology},
  volume={178},
  pages={107622},
  year={2025},
  publisher={Elsevier}
}

@techreport{gdpr2016,
  title        = {Regulation ({EU}) 2016/679 of the {European Parliament} and of the {Council} on the protection of natural persons with regard to the processing of personal data and on the free movement of such data ({General Data Protection Regulation})},
  author       = {{European Parliament and Council of the European Union, 2016}},
  institution  = {Official Journal of the European Union},
  number       = {L 119},
  pages        = {1--88},
  year         = {2016},
  note         = {OJ L 119, 4.5.2016}
}

@techreport{euaiact2024,
  title        = {Regulation ({EU}) 2024/1689 of the {European Parliament} and of the {Council} laying down harmonised rules on artificial intelligence ({Artificial Intelligence Act})},
  author       = {{European Parliament and Council of the European Union, 2024}},
  institution  = {Official Journal of the European Union},
  number       = {L 1689},
  year         = {2024},
  note         = {OJ L, 12.7.2024}
}

@techreport{dataact2023,
  title        = {Regulation ({EU}) 2023/2854 of the {European Parliament} and of the {Council} on harmonised rules on fair access to and use of data ({Data Act})},
  author       = {{European Parliament and Council of the European Union, 2023}},
  institution  = {Official Journal of the European Union},
  number       = {L 2854},
  year         = {2023},
  note         = {OJ L, 22.12.2023}
}

@techreport{dsa2022,
  title        = {Regulation ({EU}) 2022/2065 of the {European Parliament} and of the {Council} on a Single Market for Digital Services ({Digital Services Act})},
  author       = {{European Parliament and Council of the European Union, 2022}},
  institution  = {Official Journal of the European Union},
  number       = {L 277},
  pages        = {1--102},
  year         = {2022},
  note         = {OJ L 277, 27.10.2022}
}

@misc{ccpa2018,
  title        = {California Consumer Privacy Act of 2018, as amended by the {California Privacy Rights Act} of 2020},
  author       = {{State of California}},
  year         = {2020},
  note         = {Cal. Civ. Code \S\S~1798.100--1798.199.100}
}

@misc{dpdpa2023,
  title        = {The Digital Personal Data Protection Act, 2023},
  author       = {{Government of India}},
  year         = {2023},
  howpublished = {No. 22 of 2023, Gazette of India Extraordinary, Part II, Section 1},
  note         = {Received assent on 11 August 2023}
}

@incollection{lim2025regulatory,
  title={Regulatory compliance},
  author={Lim, Hannah Yee-Fen},
  booktitle={Artificial Intelligence},
  pages={149--177},
  year={2025},
  publisher={Edward Elgar Publishing}
}

@misc{noyb2024,
  author       = {{noyb -- European Center for Digital Rights}},
  title        = {{GDPR}: A Culture of Non-Compliance?},
  year         = {2024},
  howpublished = {Online},
  note         = {Available: \url{https://noyb.eu}. Accessed: May 21, 2026}
}

@article{breaux2008analyzing,
  author       = {Breaux, Travis D. and Ant{\'o}n, Annie I.},
  title        = {Analyzing Regulatory Rules for Privacy and Security Requirements},
  journal      = {IEEE Transactions on Software Engineering},
  volume       = {34},
  number       = {1},
  pages        = {5--20},
  year         = {2008},
  publisher    = {IEEE}
}

@article{sleimi2021automated,
  author       = {Sleimi, Amin and Sannier, Nicolas and Sabetzadeh, Mehrdad and Briand, Lionel C. and Ceci, Marcello and Dann, John},
  title        = {An Automated Framework for the Extraction of Semantic Legal Metadata from Legal Texts},
  journal      = {Empirical Software Engineering},
  volume       = {26},
  number       = {3},
  pages        = {43},
  year         = {2021},
  publisher    = {Springer}
}

@article{zeni2015gaiust,
  author       = {Zeni, Nicola and Kiyavitskaya, Nadzeya and Mich, Luisa and Cordy, James R. and Mylopoulos, John},
  title        = {{GaiusT}: Supporting the Extraction of Rights and Obligations for Regulatory Compliance},
  journal      = {Requirements Engineering},
  volume       = {20},
  number       = {1},
  pages        = {1--22},
  year         = {2015},
  publisher    = {Springer}
}

@inproceedings{abualhaija2025xtrareg,
  author       = {Abualhaija, Sallam and Ceci, Marcello and Sannier, Nicolas and Bianculli, Domenico and Lannier, St{\'e}phane and Siclari, Michele and Voordeckers, Olivier and Tosza, Stanis{\l}aw},
  title        = {{LLM}-assisted Extraction of Regulatory Requirements: A Case Study on the {GDPR}},
  booktitle    = {Proceedings of the 33rd IEEE International Requirements Engineering Conference (RE)},
  year         = {2025}
}

@inproceedings{kiyavitskaya2008automating,
  author       = {Kiyavitskaya, Nadzeya and Zeni, Nicola and Breaux, Travis D. and Ant{\'o}n, Annie I. and Cordy, James R. and Mich, Luisa and Mylopoulos, John},
  title        = {Automating the Extraction of Rights and Obligations for Regulatory Compliance},
  booktitle    = {Proceedings of the 27th International Conference on Conceptual Modeling (ER 2008)},
  series       = {Lecture Notes in Computer Science},
  volume       = {5231},
  pages        = {154--168},
  year         = {2008},
  publisher    = {Springer},
  address      = {Berlin, Heidelberg}
}

@inproceedings{governatori2006compliance,
  author       = {Governatori, Guido and Milosevic, Zoran and Sadiq, Shazia},
  title        = {Compliance Checking between Business Processes and Business Contracts},
  booktitle    = {Proceedings of the 10th IEEE International Enterprise Distributed Object Computing Conference (EDOC)},
  year         = {2006},
  publisher    = {IEEE}
}

@inproceedings{palmirani2018pronto,
  author       = {Palmirani, Monica and Martoni, Michele and Rossi, Arianna and Bartolini, Cesare and Robaldo, Livio},
  title        = {{PrOnto}: Privacy Ontology for Legal Reasoning about Personal Data Processing},
  booktitle    = {Proceedings of the 17th IFIP International Electronic Government Conference (EGOV)},
  year         = {2018},
  publisher    = {Springer}
}

@inproceedings{siena2008metamodel,
  author       = {Siena, Alberto and Perini, Anna and Susi, Angelo and Mylopoulos, John},
  title        = {A Meta-Model for Modelling Law-Compliant Requirements},
  booktitle    = {Proceedings of the International Workshop on Requirements Engineering and Law (RELAW)},
  year         = {2008}
}

@inproceedings{ingolfo2013nomos2,
  author       = {Ingolfo, Silvia and Siena, Alberto and Mylopoulos, John},
  title        = {Nomos 2: Making Laws Explicit in Requirements Engineering},
  booktitle    = {Proceedings of the CAiSE Forum},
  year         = {2013}
}

@inproceedings{ghanavati2014legalgrl,
  author       = {Ghanavati, Sepideh and Amyot, Daniel and Rifaut, Andr{\'e}},
  title        = {Legal Goal-Oriented Requirement Language ({Legal GRL}) for Modeling Regulations},
  booktitle    = {Proceedings of the International Workshop on Modeling in Software Engineering (MiSE)},
  year         = {2014}
}

@inproceedings{ioannidis2023gracenote,
  author       = {Ioannidis, John and Harper, James and Quah, M. S. and Hunter, D.},
  title        = {Gracenote.ai: Legal Generative {AI} for Regulatory Compliance},
  booktitle    = {Proceedings of the Legal AI and Intelligent Agents Workshop (LegalAIIA)},
  year         = {2023}
}

@article{klimas2008recitals,
  author       = {Klimas, Tadas and Vaiciukaite, Jurate},
  title        = {The Law of Recitals in {European Community} Legislation},
  journal      = {ILSA Journal of International and Comparative Law},
  volume       = {15},
  year         = {2008}
}

@misc{openai2025gpt5,
  author       = {{OpenAI}},
  title        = {Introducing {GPT-5}},
  year         = {2025},
  howpublished = {\url{https://openai.com/index/introducing-gpt-5/}},
  note         = {Accessed: 2025}
}

@article{chen2024chatgpt,
  author       = {Chen, Lingjiao and Zaharia, Matei and Zou, James},
  title        = {How Is {ChatGPT}'s Behavior Changing over Time?},
  journal      = {Harvard Data Science Review},
  volume       = {6},
  number       = {2},
  year         = {2024},
  doi          = {10.1162/99608f92.5317da47}
}

@inproceedings{dalpiaz2019requirements,
  author       = {Dalpiaz, Fabiano and Dell'Anna, Davide and Aydemir, Fatma Ba{\c{s}}ak and {\c{C}}evikol, Selin},
  title        = {Requirements Classification with Interpretable Machine Learning and Dependency Parsing},
  booktitle    = {Proceedings of the 27th IEEE International Requirements Engineering Conference (RE)},
  pages        = {142--152},
  year         = {2019},
  publisher    = {IEEE}
}

@book{cockburn2001usecases,
  author       = {Cockburn, Alistair},
  title        = {Writing Effective Use Cases},
  year         = {2001},
  publisher    = {Addison-Wesley},
  address      = {Boston, MA},
  isbn         = {0201702258},
  series       = {Agile Software Development Series}
}

@book{krippendorff2004content,
  author       = {Krippendorff, Klaus},
  title        = {Content Analysis: An Introduction to Its Methodology},
  edition      = {2nd},
  year         = {2004},
  publisher    = {Sage Publications},
  address      = {Thousand Oaks, CA}
}

@misc{tunstall2022setfit,
  author       = {Tunstall, Lewis and Reimers, Nils and Jo, Unso Eun Seo and Bates, Luke and Korat, Daniel and Wasserblat, Moshe and Pereg, Oren},
  title        = {Efficient Few-Shot Learning Without Prompts},
  year         = {2022},
  publisher    = {arXiv},
  doi          = {10.48550/arXiv.2209.11055},
  url          = {https://arxiv.org/abs/2209.11055}
}

@article{pedregosa2011scikit,
  author       = {Pedregosa, F. and Varoquaux, G. and Gramfort, A. and Michel, V. and
                  Thirion, B. and Grisel, O. and Blondel, M. and Prettenhofer, P. and
                  Weiss, R. and Dubourg, V. and Vanderplas, J. and Passos, A. and
                  Cournapeau, D. and Brucher, M. and Perrot, M. and Duchesnay, E.},
  title        = {Scikit-learn: Machine Learning in {Python}},
  journal      = {Journal of Machine Learning Research},
  volume       = {12},
  pages        = {2825--2830},
  year         = {2011}
}

@inproceedings{reimers2019sentencebert,
  author       = {Reimers, Nils and Gurevych, Iryna},
  title        = {Sentence-{BERT}: Sentence Embeddings using Siamese {BERT}-Networks},
  booktitle    = {Proceedings of the 2019 Conference on Empirical Methods in Natural Language Processing (EMNLP)},
  year         = {2019},
  publisher    = {Association for Computational Linguistics},
  url          = {https://arxiv.org/abs/1908.10084}
}

@misc{openai2025gpt54,
  author       = {{OpenAI}},
  title        = {{GPT-5.4} Model},
  year         = {2025},
  howpublished = {\url{https://developers.openai.com/api/docs/models/gpt-5.4}},
  note         = {Model snapshot: gpt-5.4-2026-03-05. Knowledge cutoff: August 2025}
}

@inproceedings{zheng2023judging,
  author       = {Zheng, Lianmin and Chiang, Wei-Lin and Sheng, Ying and Zhuang, Siyuan and
                  Wu, Zhanghao and Zhuang, Yonghao and Lin, Zi and Li, Zhuohan and
                  Li, Dacheng and Xing, Eric P. and Zhang, Hao and Gonzalez, Joseph E.
                  and Stoica, Ion},
  title        = {Judging {LLM}-as-a-Judge with {MT}-Bench and Chatbot Arena},
  booktitle    = {Advances in Neural Information Processing Systems (NeurIPS)},
  year         = {2023},
  url          = {https://arxiv.org/abs/2306.05685}
}

@article{spearman1904proof,
  author       = {Spearman, Charles},
  title        = {The Proof and Measurement of Association between Two Things},
  journal      = {The American Journal of Psychology},
  volume       = {15},
  number       = {1},
  pages        = {72--101},
  year         = {1904},
  publisher    = {University of Illinois Press}
}

@book{phoenix2024prompt,
  author    = {James Phoenix and Mike Taylor},
  title     = {Prompt Engineering for Generative AI: Future-Proof Inputs for Reliable AI Outputs},
  publisher = {O'Reilly Media},
  year      = {2024},
  isbn      = {9781098153434}
}

@article{wilcoxon1945individual,
  author  = {Frank Wilcoxon},
  title   = {Individual Comparisons by Ranking Methods},
  journal = {Biometrics Bulletin},
  volume  = {1},
  number  = {6},
  pages   = {80--83},
  year    = {1945}
}

@inproceedings{hassani2024rethinking,
  author    = {Sallam Hassani and Mehrdad Sabetzadeh and Daniel Amyot and Jia Liao},
  title     = {Rethinking Legal Compliance Automation: Opportunities with Large Language Models},
  booktitle = {Proceedings of the 32nd IEEE International Requirements Engineering Conference (RE)},
  pages     = {432--440},
  year      = {2024},
  doi       = {10.1109/RE59067.2024.00051}
}

\end{document}